\documentstyle[12pt]{article}

\def\phibar{{\bar \phi}}
\def\psibar{{\bar \psi}}

\def\lattint{\int {d^4k \over (2\pi)^4}}
\def\ktilde{\sum_\mu sin^2 k_\mu }

\begin{document}

\begin{titlepage}
\begin{flushright}
\hfil                 AZPH-TH 92-08\\
\hfil                 UCSD/PTH 92-08\\
\hfil                 BNL-47708\\
\hfil May 1992 \\
\end{flushright}
\begin{center}

{\LARGE\bf The Phase Diagram of\\ a U(1) Higgs-Yukawa Model at Finite
$\lambda$}

\vspace{0.5cm}
Anna Hasenfratz\\
Department of Physics, University of Arizona, Tucson\\
Tucson, AZ 85721, USA\\
\vspace{0.2cm}
Karl Jansen\\
Department of Physics, University of California at San Diego,\\
La Jolla, CA~92093-0319, USA  \\
\vspace{0.2cm}
and\\
\vspace{0.2cm}
Yue Shen$^{1}$\\
Physics Department, Brookhaven National Laboratory\\
Upton, NY 11973, USA \\

\vspace{0.5cm}
\end{center}

\abstract{
In this paper we investigate how
the phase diagram of a U(1)
symmetric Higgs-Yukawa system depends  on the scalar
self coupling $\lambda$.
The phase diagram of similar models with continuous symmetry were
extensively studied in the infinite scalar self coupling
$\lambda=\infty$ limit.
Recent analytical and numerical
calculations at zero  self coupling showed  qualitatively
different phase diagram,
raising the question of the $\lambda$ dependence of the phase diagram.
Here we use analytical (large $N_f$, perturbative and mean field)
approximations as well as
numerical simulations to investigate the system.
}
\vfill
\footnotetext[1]{{\em  Research supported under contract number
DE-AC02-76CH00016 with the U.S. Department of Eneregy. Accordingly,
the U.S. Government retains a non-exclusive, royalty-free license
to publish or reproduce the published form of this contribution, or
allow others to do so, for U.S. Government purposes.
}}
\end{titlepage}

\section{Introduction}

The non-perturbative studies of the Higgs-Yukawa systems were
motivated by the ever increasing top quark mass limit
and the triviality problem of the Standard Model.
The phase diagram
investigations of four dimensional Higgs-Yukawa models produced several
surprising non-perturbative  results in recent years
\cite{Goltermann}.

In the weak Yukawa coupling region a Higgs-Yukawa model with
continuous symmetry
can be in a ferromagnetic (FM) phase, a symmetric (SYM) phase
or an antiferromagnetic (AFM) phase, depending on the scalar hopping
parameter ($\kappa$) values. At large positive $\kappa$ values the continuous
symmetry is spontaneously broken and the system is in the FM phase.
The model has a massive scalar particle, one or more massless Goldstone
bosons and fermions with mass generated via the spontaneous symmetry
breaking of the scalar field. For small $\kappa$ values the system is in the
SYM phase, in which the
theory contains degenerate massive scalars and massless fermions
(assuming there are no bare fermion mass terms in the action). The FM and SYM
phases are separated by a second order phase transition line.
The critical behavior of the model along this FM-SYM phase transition line is
expected to be governed by the perturbative Gaussian
fixed point, where both the scalar and Yukawa couplings
are marginally irrelevant. In the infinite cut-off limit the
renormalized couplings vanish, $y_R=0,\lambda_R=0$. As the pure scalar model
is ``trivial", i.e. it has no other fixed point than the perturbative
one, similar behavior is expected even for strong scalar coupling as
long as the Yukawa interaction is weak.
At large negative $\kappa$ values the model is in the AFM phase. It is
generally believed that this phase is separated from the symmetric phase
by a second order phase transition line the same way as the
ferromagnetic and symmetric phases are in the positive hopping parameter
region.

For large Yukawa couplings the situation is quite different. Quenched
and unquenched  Monte Carlo simulations of the model with naive fermions,
supported by strong Yukawa coupling expansion,
revealed the existence of non-perturbative symmetric and broken phases
in the large Yukawa coupling region. These phases are separated by a
second order phase transition line where the fermions of the model have
large (at the order of the cut-off) mass, the fermions decouple in the
continuum limit leaving a non-interacting scalar theory behind.

Several numerical calculations investigated Higgs-Yukawa systems
recently \cite{Goltermann}.
Apart from studies of the $Z_2$ discrete symmetry model \cite{Lee}, all
considered the limit of infinite scalar coupling.
The simulations for both
the $U(1)$ and $SU(2)$ symmetric systems  observed the phase diagrams in
agreement with the above descriptions. With naive
fermions the phase diagrams showed perturbative
and non-perturbative (strong Yukawa coupling) SYM,
FM and AFM phases. In addition a ferrimagnetic (FI) phase was found
numerically at the intermediate Yukawa coupling
values, where both the magnetization and staggered magnetization are finite
\cite{Hasenfratz}.
All phase boundaries were claimed to be second order.
There exists a special point in the phase diagram
where three phases, symmetric, ferromagnetic and ferrimagnetic coexist.
It was speculated that this point could be a non-trivial fixed
point where the critical behavior of the system might change.
This scenario  however could not be confirmed
by Monte Carlo simulations \cite{bock2}. It was found that the critical
behavior of the system very close to this point is still consistent with the
perturbative predictions.

The vanishing scalar coupling $\lambda$ limit
received attention last year when it was
shown in the large fermion number ($N_f$) limit that the Higgs-Yukawa
model is equivalent (up to inverse cut-off corrections) to the
generalized Nambu-Jona-Lasinio type models \cite{UCSD}.
The phase diagram
was calculated  exactly in the large $N_f$ limit for
arbitrary  Yukawa  and vanishing scalar couplings. Numerical simulations agreed
surprisingly well with the large $N_f$ predictions even for $N_f=2$.
However this phase diagram turned out to be
very different from the infinite scalar coupling case.
A strong first order phase transition line was observed at $\lambda =0$.
It was found that  the FM to SYM phase
transition line, which is relevant to the Standard Model physics,
stops at the first oder phase transition line.
There is no sign for  non-trivial critical behavior along this line. All the
points on it belong to the attractive domain of the Gaussian
fixed point.
The FI phase, observed at $\lambda =\infty$ \cite{Hasenfratz},
does not exist at $\lambda =0$.

This paper is our first attempt to understand how the phase diagram and
the critical properties of the U(1) invariant Higgs-Yukawa model changes
from $\lambda =0$ to the $\lambda =\infty$ limit. In section 2 we study the
Higgs-Yukawa model under various theoretical approximations and a
typical phase diagram for small $\lambda$ is discussed. In section 3
we present our numerical results for $\lambda \le 1$ and compare them with the
theoretical expectations. In section 4 we conclude this study
and discuss the $\lambda=\infty$ limit briefly.

\section{Analytical Considerations }

\subsection{The model}

The lattice action for the $U(1)$ chiral invariant Higgs-Yukawa model
with naive fermions is defined as
\begin{equation}
S = S_f + S_H  ~.
\label{eq:action}
\end{equation}
The fermion part of the action $S_f$ is given by
\begin{equation}
S_f = \sum_{x,z} \psibar_i(x) M(x,z) \psi_i(z)~~, ~~
i = 1,2,...,N_f/2 ~~.
\label{eq:F-act}
\end{equation}
In eqn. (\ref{eq:F-act}) the fermion matrix may be written as
\begin{equation}
M(x,z) = \sum_{\mu} \gamma_\mu\left[\delta_{x+\mu,z}-\delta_{x-\mu,z}\right]
+ y\left[\phi_1(x)+i\gamma_5\phi_2(x)\right]\delta_{x,z} ~,
\label{eq:F-matrix}
\end{equation}
where $\gamma_\mu, \gamma_5$ are the Hermitian Dirac matrices and $y$ stands
for
the Yukawa coupling.
The scalar part of the action $S_H$ in eqn. (\ref{eq:action})
is given by
\begin{eqnarray}
S_H &=& -\kappa\sum_{x,\mu} \phi_a(x)\left[\phi_a(x+\mu)+\phi_a(x-\mu)\right]
+ \sum_x \phi_a^2(x)  \nonumber \\
&+&  \sum_x\lambda \left[\phi_a^2(x)-1\right]^2 ~,~~
a = 1, 2 ~~.
\label{eqnarray:H-act}
\end{eqnarray}
In the numerical simulations, we use $det M^\dagger M$ for the fermion
matrix to keep the partition function positive definite.
This is equivalent to including an extra fermion species in the action
eqn. (\ref{eq:action}).

\subsection{Large $N_f$ limit}
The model defined in eqns.(\ref{eq:action}-\ref{eqnarray:H-act})
can be solved in the large fermion number
$N_f$ limit. The $\lambda=0$ case was discussed in Ref \cite{UCSD},
and  now we include the quartic term in the analysis.

It is convenient to consider a modified form of the scalar action

\begin{eqnarray}
S_H &=& -\kappa_N \sum_{x,\mu}
\varphi_a(x)\left[\varphi_a(x+\mu)+\varphi_a(x-\mu)
\right] + \sum_x \varphi_a^2(x) \nonumber \\
&+& {\lambda_N} \left(\varphi_a^2(x)-N_f\right)^2 ~.
\label{eqnarray:F-act2}
\end{eqnarray}
The usual lattice action  eqns. (\ref{eq:action}-\ref{eqnarray:H-act})
is obtained by identifying
\begin{equation}
\kappa_N = C^2 \kappa~, \
\lambda_N = C^4 \lambda  ~,\
y_N = C y~,
\label{eq:factor}
\end{equation}
where the factor $C$ satisfies the equation
\begin{equation}
C^4 - (1-2\lambda_N N_f) C^2 - 2\lambda_N = 0
\label{eq:C}
\end{equation}
The scalar field $\varphi_a(x)$ is related to the original field $\phi_a(x)$
by
\begin{equation}
\varphi_a(x) = \phi_a(x)/C~.
\end{equation}
In the $1/N_f$ expansion  the couplings
$\tilde y_N =\sqrt {N_f}y_N$, $\tilde \lambda_N=N_f\lambda_N$
are kept fixed to be $O(1)$. As $N_f \to \infty$ the relations
in eqn. (\ref{eq:factor}) simplify, giving
\begin{equation}
\kappa = {\kappa_N \over {1-2\tilde\lambda_N }}~, \
\lambda N_f = {\tilde\lambda_N \over {(1-2\tilde\lambda_N)^2}} ~,\
y \sqrt{N_f} = { \tilde y_N \over {\sqrt{1-2\tilde\lambda_N}}} ~, \ \ \ \
{\tilde\lambda_N < {1\over 2}}~. \
\label{eq:factor2}
\end{equation}
In the large $N_f$ limit the constant mode of the scalar field dominates
the path integral which suggests the Ansatz
\begin{equation}
\varphi_1(x) = \sqrt{N_f} (a + (-1)^{\sum_\mu x_\mu} b )\;\;,
\varphi_2(x) = 0 ~,
\end{equation}
where $\sqrt{N_f}a$ and $\sqrt{N_f}b$  correspond to the
magnetization and staggered magnetization, respectively. The effective
potential at leading order is
\begin{eqnarray}
{1\over N_f} V_{eff}(a,b) &=& -8\kappa_N(a^2-b^2)+(a^2+b^2)+{\tilde\lambda_N}
\left((a^2+b^2-1)^2+4a^2b^2\right) \\ \nonumber
&-& 2 \lattint log\left[ \sum_\mu
sin^2k_\mu + \tilde y^2_N(a^2-b^2)\right] ~.
\end{eqnarray}

Minimizing $V_{eff}(a,b)$ with respect to $a,b$ can give
in general four types of solutions:

(1)Symmetric (SYM) solution: $a = 0,\  b= 0$.

(2)Ferromagnetic (FM) solution: $a \ne 0,\  b = 0$. The magnetization $a$ is
determined by the equation of state
\begin{equation}
\left.{1 \over N_f}{\partial V_{eff} \over \partial a} \right|_{b=0}=
  2a\big(1 - 8\kappa_N + 2(a^2-1){\tilde \lambda_N}
- 2\tilde y_N^2 I(\tilde y_N^2 a^2)\big)
=0~,
\label{eq:FMSYM}
\end{equation}
where the integral $I$ is defined by
\begin{equation}
I(x) = \int {d^4k\over (2\pi)^4} {1 \over \sum_\mu sin^2k_\mu + x} ~.
\end{equation}

(3)Antiferromagnetic (AFM) solution: $a = 0, \ b \ne 0$. The staggered
magnetization $b$ is given by
\begin{equation}
\left.{1 \over N_f}{\partial V_{eff} \over \partial b}\right|_{a=0}=
2b\big(1 + 8\kappa_N + 2(b^2-1)\tilde \lambda_N +
 2\tilde y_N^2 I(-\tilde y_N^2 b^2)\big)
=0~.
\label{eq:AFMSYM}
\end{equation}

(4) Ferrimagnetic (FI) solution: $a \ne 0, b \ne 0$. It is
straightforward to show that this solution can exist only when
$\tilde \lambda_N > 1/2$. However, the relation given in equation
(\ref{eq:factor2}) becomes invalid for $\tilde \lambda_N > 1/2$.
Actually it is easy to find out from equation (\ref{eq:C}) that
$\tilde \lambda_N > 1/2$  corresponds to the intermediate and strong
$\lambda$ region for the original lattice action
eqn. (\ref{eq:action}), which we will investigate
in a future study. Thus the large $N_f$ calculation indicates that
at least there is no FI phase in the weak $\lambda$ region.

In some parameter range several solutions may coexist.
It is a simple numerical exercise to evaluate the effective potential and
find the solution that gives the absolute minimum.

Although the above approach is exact in the limit $N_f \to \infty$
with fixed $\tilde\lambda_N$ and $\tilde y_N$, it only describes the small
$\lambda(=O(1/N_f))$ and $y(=O(1/\sqrt{N_f}))$
region of the original model eqn. (\ref{eq:action}). For
large $y$ values another type of large $N_f$ expansion is
possible \cite{UCSD}. We start with the action eqns. (\ref{eq:action})
and (\ref{eqnarray:F-act2}),
only now assuming that $y_N \sim O(1)$ and $\kappa_N \sim
O(1/N_f)$. After integrating out the fermions, we get an effective
action for the scalar variables
\begin{equation}
S_{eff} = S_H - {N_f\over 2} tr\log MM^\dagger ~,
\label{eq:effact}
\end{equation}
where $S_H$ and $M$ are given by eqns. (\ref{eqnarray:F-act2}) and
(\ref{eq:F-matrix}), respectively.
One may expand the
fermion determinant in powers of $1/N_f$. The leading term in $1/N_f$
fixes the amplitude of $\varphi_a(x)$
\begin{equation}
\varphi_a(x) = \varphi_0\sigma_a(x) ~, \
1 + 2{\tilde \lambda_N}({\varphi_0^2 \over N_f}-1) - {2N_f\over \varphi_0^2}
= 0 ~,
\end{equation}
where $\sigma_a(x)$ is a two-component field with unit length.
At next to leading order in $1/N_f$, the effective action
eqn. (\ref{eq:effact}) becomes, up to an additive constant, an effective
4-dimensional XY-model
\begin{equation}
S_{eff} = -\kappa_{eff}
\sum_{x,\mu}\sigma_a(x)\left[\sigma_a(x+\mu)+\sigma_a(x-\mu)\right] ~.
\end{equation}
The effective hopping parameter
$\kappa_{eff}$ is given by
\begin{equation}
\kappa_{eff} = \kappa_N\varphi_0^2+{N_f \over 2y_N^2\varphi_0^2}~.
\label{eq:XY}
\end{equation}
The XY-model is known to have second order phase transitions at
$\kappa_{eff} \approx 0.15$ (FM-SYM) and $\kappa_{eff} \approx -0.15$
(AFM-SYM).
Thus eqn. (\ref{eq:XY}) predicts the existence of two second order phase
transition lines for large $y_N$.

Combining the above two large $N_f$ expansion results,
we plot the phase diagram for  fixed ${\tilde \lambda_N}(=0.1)$ in Fig. 1.
We find the following:

(1) An FM-SYM second order phase transition line (AB).
It is described by eqn. (\ref{eq:FMSYM}) in the limit  $a\to 0$.

(2) An AFM-SYM second order phase transition line (CD).
It is given by eqn. (\ref{eq:AFMSYM}) in the limit  $b \to 0$.

(3) A first order phase transition line (EDBF). On this line
the effective potential has an AFM minimum in coexistence with either
another AFM, SYM, or FM minimum.
The staggered magnetization is discontinuous along the line DE but it is
finite on both sides of the phase transition line. The discontinuity
decreases from D to E and becomes zero at the point E.
Therefore the AFM phases in the small and large $y_N$ regions are
analytically connected.
For increasing $\lambda_N$ values the position of the DE
segment moves to larger $y_N$ and the end point E
moves to more negative $\kappa_N$ value. As $\lambda_N \to 0$ the
line CD disappears and the point E
coincides  with the point C as given in Ref \cite{UCSD}.

(4) A  second order FM-SYM phase transition line (GH) given by
eqn. (\ref{eq:XY}) setting $\kappa_{eff} = 0.15$.

(5) A second order AFM-SYM phase transition line (IJ) given by
eqn. (\ref{eq:XY}) setting $\kappa_{eff} = -0.15$.

The region between F,K,I,G is outside the validity of both large
$N_f$ expansions. The order of phase transitions for line FK,KI, and KG
can be only determined by numerical simulations.
The second order FM-SYM transition line AB ends on a first order
phase transition line but the first order line does not get critical at
this point. The whole AB line, including the endpoint B,
is in the domain of attraction of
the Gaussian fixed point.  No new non-trivial fixed point
is found in this region.

Although the above phase diagram is obtained in the large $N_f$
expansions, the bare perturbation calculation in the small $y$ region
predicts the same structure for finite $N_f$ as shown in Fig. 2a.

\subsection{Mean field calculation}

When $\lambda$ is relatively large, both large $N_f$ expansions
and the bare perturbation calculation become invalid. For a qualitative
phase diagram, one may use  mean field calculations.
Here we follow the well know
\cite{Drouffe,Smit,Bock,Lee} saddle point type mean field approximation.

Because of the $U(1)$ chiral symmetry, one may choose the Ansatz for the
saddle point
\begin{eqnarray}
\phi_1(x) &=& a + (-1)^{\sum_\mu x_\mu} b, \ \  \phi_2(x) = 0 ~, \\ \nonumber
h_1(x) &=& h + (-1)^{\sum_\mu x_\mu} h_{st}, \ \   h_2(x) = 0 ~,
\label{eq:ansatz1}
\end{eqnarray}
where $h_1(x), h_2(x)$ are the auxiliary fields introduced in the mean field
calculation. The saddle point conditions define $h$ and $h_{st}$ as
implicit functions of the magnetization $a$ and the staggered magnetization
$b$
\begin{equation}
a + {1\over 2} {u^\prime} (h + h_{st}) + {1\over 2}{u^\prime}(h-h_{st})
= 0 ~,
\label{eq:mean3}
\end{equation}
\begin{equation}
b + {1\over 2}{u^\prime}(h+h_{st}) - {1\over 2}{u^\prime}(h-h_{st})=0 ~.
\label{eq:mean4}
\end{equation}
where $u(x)$ is a function defined by
\begin{equation}
\exp\{u(x)\} = \int d\rho d\theta exp\left\{-V(\rho) - \rho x
\cos{\theta}\right\} ~,
\label{eq:u}
\end{equation}
and $V(\rho) = \rho^2 + \lambda (\rho^2-1)^2 $.

With this Ansatz, the effective potential of the system
has the form
\begin{eqnarray}
-V_{MF}(a,b) &=& 8\kappa (a^2-b^2) + 2N_f \lattint ln\left[ \sum_\mu sin^2
k_\mu + y^2(a^2-b^2)\right] \\ \nonumber
&+& h a + h_{st}b + {1\over 2} u(h+h_{st})
+ {1\over 2} u(h-h_{st}) ~.
\label{eq:mean5}
\end{eqnarray}
Minimizing $V_{MF}(a,b)$ with respect to $a,b$ gives three possible
phases

(1) SYM phase: $a=0, \ h = 0, \ b = 0, \ h_{st} = 0$.

(2) FM phase: $a \ne 0,\  h \ne 0, \ b = 0, \ h_{st} = 0$.
$a$ and $h$ satisfy the equations
\begin{equation}
a + u^\prime (h) = 0 ~,
\end{equation}
\begin{equation}
16\kappa a + 4N_f y^2a\lattint {1\over \ktilde + y^2a^2} + h = 0
{}~.
\label{eq:mean1}
\end{equation}

(3) AFM phase: $a = 0, \ h = 0, \ b \ne 0, \ h_{st} \ne 0$.
$b$ and $h_{st}$ satisfy the equations
\begin{equation}
b + u^\prime (h_{st}) = 0 ~,
\end{equation}

\begin{equation}
-16\kappa b - 4N_f y^2 b \lattint {1\over \ktilde -
y^2b^2}+h_{st} = 0 ~.
\label{eq:mean2}
\end{equation}

Using the fact that ${u^\prime}(x)$
is an odd and monotonically increasing function of $x$, one can show that
the FI ($a \ne 0, b \ne 0$) solution is excluded in the mean field
approximation.

Similar to the large $N_f$ calculation, in some parameter range several
solutions may coexist.
Again one needs to evaluate the effective potential and
find the solution that gives the absolute minimum.

The above mean field calculation breaks down
in the large $y$ region due to the following simple reasoning:
When $y$ is large, the fermion determinant can be expanded and to leading
order in $1/y^2$ we have an effective action
\begin{eqnarray}
S[\phi] &=& -\kappa \sum_{x,\mu}\left [ \rho(x)\rho(x+\mu)\sigma_a(x)
\sigma_a(x+\mu) + (\mu \to -\mu) \right]  \\ \nonumber
&+& \sum_x \left\{ \rho(x)^2 + \lambda [\rho(x)^2-1]^2 -
2N_f\log\rho^2(x)\right\}~,
\end{eqnarray}
where the radial and angular notation for the $\phi$ field is used
\begin{eqnarray}
\phi_1(x) &=& \rho(x)\sigma_1(x),  \ \sigma_1(x) = cos(\theta(x)) ~,\\
\nonumber
\phi_2(x) &=& \rho(x)\sigma_2(x),  \ \sigma_2(x) = sin(\theta(x))~.
\end{eqnarray}
If we stay on the $\kappa=0$ axis,
the system becomes a collection of uncorrelated
rotators and the U(1) chiral symmetry will be unbroken. An expansion around
the $\kappa=0$ axis will have a finite radius of convergence. Thus by
analytical continuation we expect the system to be in the symmetric
phase in the large $y$
region around $\kappa=0$. However, this symmetric phase in the large
$y$ region is not predicted by the mean field calculation.

We comment here that for the Higgs-Yukawa model the saddle point
type mean field approximation is not equivalent to the variational type
mean field approximation. With the saddle point type approximation, one
is not guaranteed to get an upper bound on the free energy. If the fluctuation
around the saddle point is large, the true free energy may be completely
different from the saddle point estimate. In contrast, the variational
type mean field approximation gives the rigorous upper bound of the free
energy.
Unfortunately, the variational calculation can not be completed without
further approximation (small or large $y$ expansion \cite{Tsypin}) for
the Higgs-Yukawa model.

A different type of mean field approximation can be performed in the large $y$
region. If we expand the fermion determinant to next leading order in
$1/y^2$, the action becomes
\begin{eqnarray}
S &=& - \sum_{x,\mu}\left\{ \left[\kappa\rho(x)\rho(x+\mu)+{N_f\over
2 y^2\rho(x)\rho(x+\mu)}\right]\sigma_a(x)\sigma_a(x+\mu)
+ (\mu \to -\mu)\right\} \\ \nonumber
&+& \sum_x \left\{ \rho(x)^2 + \lambda [\rho(x)^2-1]^2 -
2N_f\log\rho^2(x)\right\}~.
\label{eqnarray:Largey}
\end{eqnarray}
The radial interaction part may be approximated by its mean field value
\begin{equation}
<\rho> = {\int^\infty_0 d\rho \rho^2 e^{-[\rho^2+\lambda(\rho^2-1)^2-2N_f
\log\rho^2]} \over
\int^\infty_0 d\rho \rho e^{-[\rho^2+\lambda(\rho^2-1)^2-2N_f
\log\rho^2]} } ~,
\end{equation}
and the action eqn. (30) becomes an effective action for the
XY-model
\begin{equation}
S_{eff} = -\kappa_{eff} \sum_{x,\mu} \sigma_a(x)\left[\sigma_a(x+\mu)
+ \sigma_a(x-\mu)\right] ~,
\end{equation}
where the effective hopping parameter $\kappa_{eff}$ is given by
\begin{equation}
\kappa_{eff} = \kappa <\rho>^2 + {N_f\over 2y^2}<{1\over \rho}>^2~.
\end{equation}
This approximation can be justified as a leading order $1/y^2$
expansion with the assumption that $\kappa \sim O(1/y^2)$.
The XY-model has second order phase transitions at $\kappa_{eff} \approx
0.15$ (FM-SYM) and $\kappa_{eff} \approx -0.15$ (AFM-SYM). Thus for the
original
Higgs-Yukawa model we get the second order phase transition lines at
\begin{equation}
\kappa <\rho>^2 + {N_f\over 2y^2}<{1\over \rho}>^2 \approx \pm 0.15~.
\end{equation}

The results of this mean field calculation are plotted in Fig. 2a and 2b in the
large $y$ region. We want to mention that for
small $\lambda$ and $y$ the mean field results are compatible with the bare
perturbation calculations which are plotted in Fig. 2a.

At $\lambda =1$ both large $N_f$ approximation and perturbation theory
break down.  Fig. 2b shows the phase diagram predicted by the weak and strong
$y$ mean field calculation  at $\lambda=1$.

\section{Numerical results}

We have performed  numerical simulations at $\lambda = 0.0156, N_f =
2,\ 10$ and  $\lambda= 0.1,\  1.0$, $N_f=2$.
The Hybrid Monte Carlo method \cite{Kennedy} was used for the dynamical fermion
simulations. Each molecular
dynamics trajectory consists of $10$ steps with step size chosen such that
the acceptance rate is around $80\%$.
To decide the order of the phase transition,
we looked for  hysteresis effects in the thermocycles.
For each data point
in the thermocycle about 20
trajectories are used as warmup and 100-200 trajectories are used
in the measurement.

The magnetization $v$ is defined as
\begin{equation}
v = <\sqrt{\phibar_a^2}>, \ \   \phibar_a = {1\over L^4}\sum_x \phi_a(x) ~,
\end{equation}
and the staggered magnetization $v_{st}$ is
\begin{equation}
v_{st} = <\sqrt{\phibar^2_{st,a}}>, \ \  \phibar_{st,a} = {1\over L^4} \sum_x
(-1)^{\sum_\mu x_\mu} \phi_a(x) ~,
\end{equation}
where $L$ is the linear size of the lattice.
$v$ and $v_{st}$ are measured and used as the order parameters to
determine the phase.
The measurements are done on $4^4$ lattices.
A $4^4$ lattice is certainly not sufficient to distinguish second order
and weakly first order transitions. However, the
combination and the agreement
of the numerical and analytical results allow us to
determine the order of the phase transitions reliably.

At $\lambda = 0.0156$, we have simulation results for both $N_f = 2$ and
$10$. For $N_f = 10$, the data can be compared directly with the
large $N_f$ calculation. We find complete agreement with the theoretical
predictions as shown in Fig. 1.
Along the line of the first order phase transition where the
hysteresis effects are very strong, it is not easy to determine exactly
the position of the phase transition. But the amplitude of $v$ and
$v_{st}$ agree very well with the large $N_f$ prediction.
Thus the phase transition line predicted by the large $N_f$ expansion
should be reliable.
For  large $y$, the SYM region is quite narrow
for $N_f=10$, making it difficult to establish this region numerically.
However, the existing results \cite{UCSD} at
$\lambda=0$ and our simulation results presented in the next paragraph
at $N_f = 2$ should be sufficiently convincing that the SYM phase indeed
exist in this region.

The phase diagram for $\lambda=0.0156, N_f = 2$ is plotted in Fig. 2a.
The transition points agree well with both the bare perturbation calculation
and the large $N_f$ expansion in the small $y$ region.
The agreement with the large $N_f$ calculation is
probably due to the fact that the
effective fermion flavor number around second order phase transition
lines is $32$ because of the lattice doubling effect.
Up to $\lambda = 1$ the picture remains qualitatively the same.
The phase diagram for
$\lambda = 1, N_f = 2$ is shown in Fig. 2b.
As $\lambda = 1$ is already a coupling of intermediate strength,
neither bare perturbation theory nor the large $N_f$
approximations agree with the data points.
However, the mean field
approximation -from which the curves are plotted in the figure-
shows reasonable agreement with the simulation result.
In particular, there is no indication for an FI phase  or a non-trivial
fixed point at any of these coupling values.

In Fig. 3 we give examples for thermocycles along
the first order phase transition line EDBF in Fig. 2 (see Fig. 1 for
notations) for $\lambda=0.0156,
N_f=2$. The data is compared to the
bare perturbation theory (dotted lines). Fig. 3a corresponds to an
AFM-AFM \footnote{Note that this figure also contains the
continuous phase transition from the
symmetric to the antiferromagnetic phase at
$\kappa \approx -0.16$}
transition, Fig. 3b to a SYM-AFM and Fig. 3c to an FM-AFM transition.
In all cases $v_{st}$ is plotted.
All thermocycles show hysteresis effects in
agreement with our theoretical expectations.
There is no sign of the first order line becoming critical at the end
point B of the FM-SYM second order phase transition line.

\section{Conclusion and discussions}

In this paper we have investigated the phase diagram of a
U(1)$\otimes$U(1) Higgs-Yukawa model with fluctuating length
Higgs field. The results obtained by various analytical methods and
numerical simulations show, up to the Higgs self coupling
$\lambda=1$, a phase diagram as summarized in
Fig. 1. This picture is quite different from the published results at
$\lambda=\infty$. We observe strong first order phase transitions from
the antiferromagnetic phase to the ferromagnetic, symmetric and
antiferromagnetic phases. The second order
line between the perturbative ferromagnetic and the symmetric phases,
that can be relevant for the Standard Model, ends on this first order
phase transition line. At this end point
the first order line does not become critical,
no new non-trivial critical behavior is expected,
the whole second order line is
in the domain of attraction of the Gaussian fixed point.
In addition, we do not find a ferrimagnetic phase up to $\lambda=1$.

In order to understand the critical behavior of the Higgs-Yukawa model
one has to find out if the above  picture for the  phase diagram
remains the same as one increases $\lambda$ further. If this is indeed
the case, it would provide a natural explanation of why the
critical indices do not change along the second order  line at
$\lambda=\infty$\cite{Bock}.
Another possibility would be that the phase diagram changes qualitatively
at some strong $\lambda$ value and eventually merges into the published
result \cite{Hasenfratz} at $\lambda=\infty$ .
The mean field picture given in section 2 remains unchanged
up to $\lambda=\infty$. However,
it is unclear how reliable these approximations really are.
To determine the critical properties of the large and infinite $\lambda$
systems requires much more computer time, or different analytical
approaches. This work is in progress and the result
will be reported in a future publication.

\noindent{\bf Acknowledgement}

We thank Peter Hasenfratz for useful discussions. Much of the work
has been done during the period when Y. Shen was visiting UCSD.
He thanks the High Energy Theory group at UC San Diego for its hospitality.
Our work is supported by DOE grants at UC San Diego (DE-FG-03-90ER40546)
and Brookhaven National Laboratory (DE-AC02-76CH00016). The numerical
simulations are performed at UCSD Supercomputer Center.

\pagebreak

\pagebreak

\begin{figure}
\vspace{1.0cm}
\caption{The phase diagram in the large $N_f$ limit at
$\tilde\lambda_N = 0.1$ and $N_f = 10$. The MC data are indicated by  circles
where the solid symbol denotes second and the open symbols first order
phase transitions. The solid and dashed lines are the
results from the $1/N_f$ expansions, where the solid lines represent
second order and the dashed line first order phase transitions. In the
middle of the phase diagram, where the $1/N_f$ expansions break down,
the dotted lines indicate how the phase transition might continue.}
\end{figure}

\begin{figure}
\vspace{1.0cm}
\caption{Phase diagram at $\lambda = 0.0156$ (a) and $\lambda=1.0$ (b),
both with
$N_f = 2$.
Here as in Fig.1 solid symbols denote second and open symbols first
order phase transitions. In the small $y$
region the solid and dashed lines are obtained by (a) bare perturbation
calculation; (b) mean field calculation.
In both (a) and (b) the lines in the
large $y$ region are obtained from the mean field theory given in the
second part of the section 2.3. Solid lines represent second order and
dashed lines first order phase transitions. The dotted lines only
indicate a possibility how the phase transition lines may continue.}
\end{figure}

\begin{figure}
\vspace{1.0cm}
\caption{Hysteresis effects for the staggered magnetization $v_{st}$
are shown along the first order phase
transition line EDBF (see Fig.1). The data are taken at $\lambda=0.0156$
and $N_f=2$. The solid symbols represent the first half of the
thermocycle and the open ones the way back. The solid lines are only
connecting the data points to guide the eye. The dotted lines are the
results from perturbation theory.
(a) a point taken between D and E (AFM-AFM) (Note that in addition to
the AFM-AFM phase transition also the SYM-AFM phase transition at
$\kappa \approx -0.16$ is shown.) (b) a point
between D and B (SYM-AFM). (c) a point taken between B and F (FM-AFM).
The discontinuity of the
hysteresis loop becomes smaller closer to point E indicating a weaker
first order phase transition.}
\end{figure}

\end{document}